\documentclass[trackchanges, twocolumn]{aastex701}
\usepackage{comment}
\usepackage{amsmath}

\begin{document}

\linenumbers

\title{Successive Partial Disruptions with Orbital Precession in a White Dwarf--Black Hole System \\for Repeating GRB~250702B}

\author[orcid=0000-0003-2477-9146]{Yuri Sato}
\affiliation{Astronomical Institute, Graduate School of Science, Tohoku University, Sendai 980-8578, Japan}
\email[show]{yuri@astr.tohoku.ac.jp}  

\author[orcid=0009-0000-1257-5133]{Rin Oikawa}
\affiliation{Astronomical Institute, Graduate School of Science, Tohoku University, Sendai 980-8578, Japan}
\email{}

\author[]{Kazuma Kato}
\affiliation{Astronomical Institute, Graduate School of Science, Tohoku University, Sendai 980-8578, Japan}
\email{}

\author[orcid=0000-0002-9350-6793]{Tatsuya Matsumoto}
\affiliation{Department of Astronomy, School of Science, The University of Tokyo, 7-3-1 Hongo, Bunkyo-ku, Tokyo 113-0033, Japan}
\email{}

\author[orcid=0000-0003-4299-8799]{Kazumi Kashiyama}
\affiliation{Astronomical Institute, Graduate School of Science, Tohoku University, Sendai 980-8578, Japan}
\email{}

\begin{abstract}
The peculiar gamma-ray burst GRB~250702B is the longest event ever observed, lasting about one day and exhibiting four prompt-emission flares of $\sim100$~s with irregular recurrence intervals of at least one hour. To explain this hierarchy of timescales, we consider a scenario in which a stellar object undergoes repeated partial tidal disruptions by a black hole (BH).
We find that if a white dwarf (WD) is on a highly eccentric orbit ($e\approx0.97$) around an intermediate-mass BH with
$M_{\rm BH}\lesssim10^{6}\,M_\odot$ and
$a = 60\,R_\odot\left(M_{\rm BH}/10^{6}\,M_\odot\right)^{1/3}$,
the observed properties of GRB~250702B can be reproduced. In this framework, the duration of each flare is determined by the viscous accretion timescale of material stripped near pericenter, with a typical mass $\Delta M \approx 2\times10^{-2}\,M_\odot$. The minimum recurrence time corresponds to the orbital period, while the total activity period is set by the secular orbital evolution timescale leading to the complete disruption of the WD.
Furthermore, if $M_{\rm BH}\gtrsim10^{5}\,M_\odot$ and the orbit has a minimum polar angle relative to the BH equatorial plane of $\theta_{\rm min}\gtrsim0.12~{\rm rad}$, relativistic frame dragging induces $\gtrsim0.1$~rad precession of the orbital angular momentum between successive pericenter passages, comparable to a typical GRB jet half-opening angle, resulting in intermittent alignment with the observer and irregular flare spacing.
The WD experiences $\approx40$ jet-launch episodes before complete disruption, but only four are expected to be observed on-axis. The remaining off-axis jets become visible at late times, enhancing the radio afterglow by about an order of magnitude, providing a testable prediction of this scenario.
\end{abstract}


\section{Introduction}

Recently, an exceptionally long-duration gamma-ray burst, GRB~250702B, was detected, exhibiting the longest prompt emission duration ever observed \citep{Neights2025}.
The event was first discovered by the {\it Einstein Probe} (EP) in X-rays \citep{EP2025} and was subsequently detected by the {\it Fermi} Gamma-ray Burst Monitor ({\it Fermi}/GBM) \citep{Neights2025}.
The first gamma-ray flare occurred approximately 10~hours after the initial X-ray detection, followed by several subsequent flares with irregular separations ranging from $\sim$1 to $\sim$25~hours \citep{EP2025,Zhang2025,Oganesyan2025,Neights2025}.
Each flare lasts for $\delta t_\mathrm{flare,obs}\sim100$~s.
The total activity duration reached $\delta t_\mathrm{tot,obs}\sim1.6$~d, with an isotropic-equivalent luminosity of $L_{\gamma,\mathrm{iso,obs}}\approx5\times10^{51}$~erg~s$^{-1}$ at $z=1.036$ \citep{JWST2025}.
Multiwavelength follow-up observations detected X-ray, infrared, and radio emission \citep{Levan2025,Carney2025,OConnor2025}, and the transient was localized at an offset of 5.7~kpc from the center of its host galaxy \citep{Carney2025}.

The coexistence of short ($\sim100$~s) flares, irregular hour-scale recurrence, and prolonged day-scale activity makes GRB~250702B highly unusual among known GRBs, suggesting a physical origin distinct from both compact-object mergers and collapsars.
Such hierarchical timescales have been proposed to arise in tidal disruption events (TDEs) \citep{Levan2025,Beniamini2025,Eyles-Ferris2025,EP2025,Granot2025,Yuan2026}.

A similar phenomenology was observed in the first known jetted TDE, Swift~J1644+57, which showed multiple high-energy flares separated by $\sim$day-long intervals, while individual flares rose on $\sim100$~s timescales \citep{Burrows2011,Levan2011}.
These timescales have been interpreted in a white dwarf--intermediate-mass black hole (WD--IMBH) disruption scenario, in which the short timescale reflects the accretion time, and the longer one corresponds to the orbital period of the WD \citep{Krolik2011}. 
In contrast, GRB~250702B exhibits flares recurring at irregular intervals as short as $\Delta t_\mathrm{flare,min,obs}\sim1$~hour in the observer frame.
This hierarchy points to a more compact system in which the WD survives multiple close encounters with the BH, as expected for partial tidal disruptions \citep[e.g.,][]{Zalamea2010,MacLeod2014,Chen2023,Lau2025,Chen2025}.

In this Letter, we propose that the temporal hierarchy of GRB~250702B arises from successive partial tidal disruptions of the WD orbiting the BH.
If the WD approaches sufficiently close to the BH, frame dragging induces relativistic precession of its orbit.
At each pericenter passage, a fraction of the WD is stripped and circularizes into a transient accretion disk, launching a relativistic jet.
Orbital precession causes the jet-launching direction to vary between passages, so that most jets are viewed off-axis, with only a few aligning with the observer’s line of sight and producing detectable prompt gamma-ray flares.
This framework will explain the coexistence of short-duration flares, irregular recurrence, and prolonged activity in GRB~250702B, and predicts enhanced late-time radio emission from the cumulative contribution of multiple off-axis jets.

\section{Physical Picture and Timescale Hierarchy}
\label{sec:model}

\begin{figure*}
\vspace*{5pt}
\centering
\includegraphics[width=0.95\textwidth]{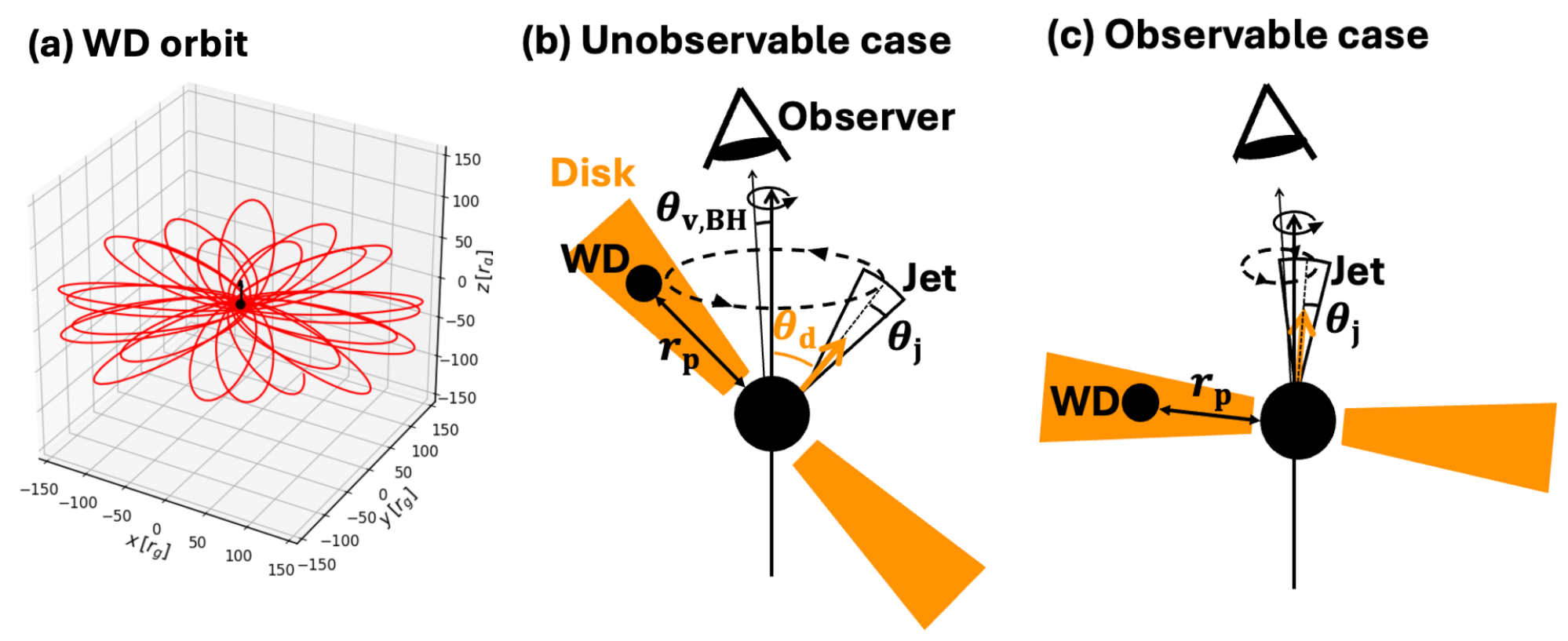}
\vspace{-0.3cm}
\caption{
Our physical picture for GRB~250702B.
Panel (a) shows the geodesic orbit of the WD, where the red line represents our numerical result with $a = 30\,R_\odot$, $e = 0.97$, $M_{\rm BH} = 1\times10^5\,M_\odot$, $a_{\rm spin}=0.9$, and $\theta_{\rm min} = 0.12~\mathrm{rad}$ ($7^\circ$).
The BH spin axis is indicated by the vertical black line.
The orbit exhibits relativistic precession.
Mass loss from the WD is not included in this calculation.
Panel (b) illustrates an unobservable configuration.
An accretion disk forms with its angular-momentum vector tilted by an angle $\theta_{\rm d}$ with respect to the BH spin axis.
As a result, the jet undergoes Lense--Thirring precession.
If $|\theta_{\rm v,BH} - \theta_{\rm d}| > \theta_{\rm j}$, the jet does not intersect the line of sight, and the prompt emission remains unobservable.
Panel (c) shows an observable configuration.
If $|\theta_{\rm v,BH} - \theta_{\rm d}| \lesssim \theta_{\rm j}$, the jet remains continuously observable throughout the Lense--Thirring precession.}
\label{fig:picture}
\end{figure*}

\subsection{Repeating Partial Disruptions}
\label{sec:pTDE}

We consider a scenario in which a WD of mass $M_{\rm WD}$ and radius $R_{\rm WD}$ is captured by a BH of mass $M_{\rm BH}$ and dimensionless spin parameter $a_{\rm spin}$, most likely as a result of dynamical scattering within its host stellar system.
After capture, the WD settles onto an eccentric bound orbit and undergoes a sequence of partial tidal disruption events.
During each pericenter passage, a fraction of the WD mass is stripped, remains gravitationally bound to the BH, and is ultimately accreted, while the stellar core survives and continues on a bound orbit.
Such repeating partial tidal disruptions have been demonstrated in previous studies \citep[e.g.,][]{Zalamea2010, MacLeod2014, Chen2023, Chen2025}.

The mass lost during a single pericenter passage can be estimated following
\citet{Zalamea2010} as
\begin{equation}
   \Delta M(\beta) \approx 6.1\left(1 - \frac{M_{\rm WD}}{M_{\rm Ch}}\right)^{0.67}\left(1 - \frac{1}{2\beta}\right)^{5/2}M_{\rm WD},
\label{eq:Delta_M}
\end{equation}
where $M_{\rm Ch} = 1.43\,M_\odot$ is the Chandrasekhar mass, $\beta \equiv r_{\rm T}/r_{\rm p}$ is the penetration factor, $r_{\rm T}\equiv R_{\rm WD}(M_{\rm BH}/M_{\rm WD})^{1/3}$ is the tidal radius, and $r_{\rm p}$ is the pericenter distance.
For such repeated partial tidal disruptions to occur, the encounter must be sufficiently weak, requiring
\begin{equation}
\beta < 1 .
\label{eq:beta}
\end{equation}
In addition, the pericenter distance must lie outside the BH horizon in order to avoid a direct relativistic plunge,
\begin{equation}
\label{eq:pTDE}
r_{\rm p} > r_{\rm h},
\end{equation}
where $r_{\rm h} \equiv r_{\rm g}\left(1+\sqrt{1-a_{\rm spin}^2}\right)$ is the radius of the event horizon of a Kerr BH with $r_{\rm g}=GM_{\rm BH}/c^2$.
Under these conditions, the stellar core remains gravitationally bound to the BH after each passage and continues on a stable eccentric orbit.
As a result, multiple partial tidal disruption events can occur before the WD is fully disrupted.

In our picture, as the WD loses mass, its radius increases due to the mass–radius relation \citep[see][]{Zalamea2010}, and the expansion of the WD can dominate over the evolution of its Roche lobe, thereby enhancing the degree of Roche lobe overflow and ultimately leading to full disruption.
The resulting orbital evolution is assumed to be governed primarily by the secular timescale.

Within this framework, we associate the observed three hierarchical timescales, $\sim 100$~s, $\sim 1$~h, and $\sim 1.6$~d, with the viscous timescale, the orbital period, and the secular evolution timescale, respectively. 
In the regime $t_{\rm vis} \ll P_{\rm orb}$, angular momentum can in principle be efficiently redistributed through the accretion disk. 
In the case of nearly conservative mass transfer, this leads to orbital expansion and suppression of further tidal stripping \citep[e.g.,][]{King2022,Eyles-Ferris2025}. 
On the other hand, if a fraction of the transferred mass is lost from the system via disk winds and/or jets, the evolution becomes non-conservative, allowing for net angular momentum loss. 
In our picture, a significant fraction of the transferred mass is accreted onto the BH, while the remaining fraction is expelled via outflows, carrying away a substantial amount of angular momentum. 
This leads to a non-conservative evolution of the system. 
In such a regime, the expansion of the WD exceeds that of the Roche-lobe, resulting in progressively enhanced mass stripping\footnote{We parameterize $(\dot{J}/J) \sim f (\dot{M}_{\rm WD}/M_{\rm WD})$ and use Eq.~(12) of \citet{King2022} to determine the condition on $f$ for which the WD expansion rate exceeds that of the Roche-lobe, requiring:
$\frac{\dot{R}_{\rm L}}{R_{\rm L}} - \frac{\dot{R}_{\rm WD}}{R_{\rm WD}} 
= -2 \frac{\dot{M}_{\rm WD}}{M_{\rm WD}} 
\left( \frac{5}{6} + \frac{\zeta}{2} - \frac{M_{\rm WD}}{M_{\rm BH}} \right) 
+ 2 \frac{\dot{J}}{J} - \frac{\dot{e}}{1+e}<0$.
Here, the parameter $\zeta$ is defined by $\dot{R}_{\rm WD}/R_{\rm WD} = \zeta \dot{M}_{\rm WD}/M_{\rm WD}$, 
and we adopt $\zeta \simeq -1.4$ for a $M_{\rm WD} \sim 1\,M_\odot$ WD. 
Substituting our parameters determined in Section~\ref{sec:parameter}, 
together with $\dot{M}_{\rm WD} \sim -\frac{\Delta M}{P_{\rm orb}}$ and $\dot{e} \sim -\frac{e}{t_{\rm sec}}$, 
we obtain the condition $f > 0.45$, indicating that angular momentum removal via outflows is required for a non-conservative evolution.
In this regime, the relative expansion rate between the Roche-lobe radius and the WD radius evolves on the secular timescale (Eq.~(\ref{eq:delta_t_tot})).}.
In the following, we focus on the system in this non-conservative regime.

\subsection{Three Characteristic Timescales}
\label{sec:timescale}
\subsubsection{Flare duration: viscous timescale}
\label{sec:100s}

At each pericenter passage, the stripped material is assumed to circularize on a timescale much shorter than the orbital period, forming an accretion disk around the BH.
We therefore identify the duration of an individual gamma-ray flare with the viscous timescale of this disk.
The circularization radius is taken to be $r_{\rm d} \sim 2 r_{\rm p} = 2a(1-e)$, where $a$ is the semi-major axis and $e$ is the eccentricity.
The viscous timescale can be estimated as
\begin{eqnarray}
    \delta t_\mathrm{flare,obs} &\approx& (1+z)t_{\rm vis} \nonumber\\
    &=&\frac{1+z}{h^2\alpha}\sqrt{\frac{r_{\rm d}^3}{GM_{\rm BH}}}
    \sim \frac{1+z}{h^2\alpha}\sqrt{\frac{8R_{\rm WD}^3}{GM_{\rm WD}\beta^3}} \nonumber\\
    &\sim& 100~\left(\frac{1+z}{2}\right)\left(\frac{\beta}{0.6}\right)^{-\frac{3}{2}}
    \left(\frac{h}{0.5}\right)^{-2}
    \left(\frac{\alpha}{0.3}\right)^{-1}\nonumber\\
    &&\times\left(\frac{M_{\rm WD}}{1~M_\odot}\right)^{-\frac{1}{2}}
    \left(\frac{R_{\rm WD}}{0.01~R_\odot}\right)^{\frac{3}{2}}~\rm{s},
    \label{eq:delta_t_flare}
\end{eqnarray}
where $h$ is the disk aspect ratio, $\alpha$ is the viscosity parameter, and $G$ is the gravitational constant.
In the regime considered here, the accretion rate is super-Eddington, which implies an increased aspect ratio \citep[e.g.,][]{Abramowicz1988,Sadowski2015}. 
We adopt a representative value $h = 0.5$. 
We further use $\alpha = 0.3$, consistent with observationally inferred values for hot, fully ionized accretion flows \citep{Martin2019}.
With these parameters, Eq.~(\ref{eq:delta_t_flare}) implies that achieving the flare duration of $\sim100~\rm s$ requires the disruption of a compact object, such as the WD.

\subsubsection{Minimum flare interval: orbital period}
\label{sec:1hour}

Partial tidal disruptions occur at each pericenter passage \citep[e.g.,][]{Zalamea2010, MacLeod2014, Chen2023, Chen2025}.
The minimum flare interval corresponds to the orbital period,
\footnotesize
\begin{eqnarray}
    \Delta t_{\rm flare,min,obs} &\simeq& (1+z)P_{\rm orb}
    = 2\pi(1+z)\sqrt{\frac{a^{3}}{G M_{\rm BH}}}\nonumber\\
    &\sim& 1~\left(\frac{1+z}{2}\right)
    \left(\frac{a}{30~R_\odot}\right)^{\frac{3}{2}}
    \left(\frac{M_{\rm BH}}{1\times10^{5}~M_\odot}\right)^{-\frac{1}{2}}
    ~{\rm h}.\nonumber \\
    \label{eq:Delta_t_flare_min}
\end{eqnarray}
\normalsize
To ensure that the WD can undergo many partial disruption events without the orbit shrinking significantly, the orbital evolution driven by gravitational-wave emission must be slow compared to the orbital period:
\begin{equation}
    P_{\rm orb} \ll \left| \frac{e}{\dot e} \right|_{\rm GW}.
    \label{eq:gw_condition}
\end{equation}

\subsubsection{Total activity duration: secular evolution}
\label{sec:1day}

As the WD loses mass during each pericenter passage, the orbit evolves on a secular timescale.
The corresponding secular evolution timescale can be estimated as
\footnotesize
\begin{eqnarray}\label{eq:delta_t_tot}
\delta t_\mathrm{tot,obs} &\approx& (1+z)t_{\rm sec}
    \sim (1+z)\frac{M_{\rm WD}}{\Delta M} P_{\rm orb}\nonumber\\
&\sim& 1.6~\left(\frac{1+z}{2}\right)
\left(\frac{\Delta t_{\rm flare,min,obs}}{1~{\rm h}}\right)
\left(\frac{\Delta M}{2\times10^{-2}~M_\odot}\right)^{-1}
\nonumber\\
    &&\times\left(\frac{M_{\rm WD}}{1~M_\odot}\right)
~{\rm d}. 
\end{eqnarray}
\normalsize
%


\subsection{Parameter Constraints from Observed Timescales}
\label{sec:parameter}

By combining Eqs.~(\ref{eq:delta_t_flare}), (\ref{eq:Delta_t_flare_min}), and (\ref{eq:delta_t_tot}) with the observed timescale of GRB~250702B, and adopting typical WD parameters of $M_{\rm WD}\sim 1\,M_\odot$ and $R_{\rm WD}\sim 0.01\,R_\odot$, together with accretion-disk parameters of $h \sim 0.5$ and $\alpha \sim 0.3$, we obtain 
\footnotesize
\begin{align}
&\beta \approx 0.6\left(\cfrac{\delta t_{\rm flare,obs}}{100~{\rm s}}\right)^{-\frac{2}{3}}, \label{eq:result_beta}\\
&\left(\frac{a}{R_\odot}\right)^3 \left(\frac{M_{\rm BH}}{M_\odot}\right)^{-1} \approx 0.2 \left(\cfrac{\Delta t_{\rm flare,min,obs}}{1~{\rm h}}\right)^{2},\label{eq:result:a-M}\\
&\Delta M \approx 2\times 10^{-2}\,M_\odot\left(\cfrac{\Delta t_{\rm flare,min,obs}}{1~{\rm h}}\right)\left(\cfrac{\delta t_{\rm tot,obs}}{1.6~{\rm d}}\right)^{-1}.\label{eq:result_dM}
\end{align}
\normalsize
The inferred value of $\beta$ satisfies Eq.~(\ref{eq:beta}).
At this stage, the combination $a^3/M_{\rm BH}$ is constrained.
Using these constraints, the eccentricity can be estimated from
\begin{eqnarray}
1-e &=& \frac{r_{\rm p}}{a}=\frac{r_{\rm T}}{a\beta}\nonumber \\
&\approx& 0.03 \left(\cfrac{\delta t_{\rm flare,obs}}{100~{\rm s}}\right)^{\frac{2}{3}}\left(\cfrac{\Delta t_{\rm flare,min,obs}}{1~{\rm h}}\right)^{-\frac{2}{3}},\label{eq:result_e}
\end{eqnarray}
yielding a highly eccentric orbit with $e \approx 0.97$.
Moreover, Eq.~(\ref{eq:Delta_M}) gives $\Delta M(\beta \approx 0.6) \approx 2\times 10^{-2}\,M_\odot$, consistent with Eq.~(\ref{eq:result_dM}).
Finally, the orbital evolution driven by gravitational radiation satisfies Eq.~(\ref{eq:gw_condition}), since the characteristic timescale $|e/\dot{e}|_{\rm GW} \sim 100~\mathrm{yr}$ is much longer than the orbital period.

Using the parameters inferred above, we verify whether the system terminates after a finite number of partial disruption events.
The mass-loss episodes are expected to occur periodically $\approx 1.6~{\rm d}/1~{\rm h}\approx40$ times.
After $\sim 40$ passages, the WD mass is expected to decrease to of order $\sim 0.1\,M_\odot$.
Using the mass--radius relation of \citet{Zalamea2010},
$R_{\rm WD} = 0.013\,R_\odot 
\left(\frac{M_{\rm Ch}}{M_{\rm WD}}\right)^{1/3}
\left(1 - \frac{M_{\rm WD}}{M_{\rm Ch}}\right)^{0.447}
\sim 0.02\,R_\odot$,
the corresponding tidal radius at the 40th orbit becomes  
$r_{\rm T} \sim 0.03\,R_\odot \left(\frac{M_{\rm BH}}{M_\odot}\right)^{1/3}$.
Using the orbital parameters inferred from the observed timescales, $(a/R_\odot)^3 (M_{\rm BH}/M_\odot)^{-1} \sim 0.1$, the pericenter distance is $r_{\rm p} \sim 0.02\,R_\odot \left(\frac{M_{\rm BH}}{M_\odot}\right)^{1/3}$.
Since $r_{\rm T} > r_{\rm p}$ at this stage, the WD would undergo complete tidal disruption on the $\sim 40$th passage.

\subsection{Jet Production and Luminosity}
\label{sec:jet}

During each pericenter passage, a fraction of the WD mass is stripped. 
A portion of this material is accreted onto the BH through an accretion disk, while the rest is lost from the system (e.g., via outflows).
We assume that the accretion process launches a relativistic jet powered by the Blandford--Znajek (BZ) mechanism, which extracts rotational energy from the spinning BH via magnetic fields \citep{Blandford1977}.
Assuming that the jet has a half-opening angle $\theta_{\rm j}$ and that the emission is approximately uniform within the jet cone, the isotropic-equivalent gamma-ray luminosity is given by
\footnotesize
\begin{eqnarray}
    L_{\gamma, \mathrm{iso,obs}} &\approx& \frac{2\eta_{\rm BZ}\eta_{\rm rad}(1-f)\,\Delta M\,c^2}{\theta_{\rm j}^2\,\delta t_\mathrm{flare,obs}}\nonumber\\ 
    &\sim& 5\times10^{51}~\eta_{\rm BZ}\left(\frac{\eta_{\rm rad}}{0.1}\right) \left(\frac{\delta t_{\rm flare,obs}}{100~{\rm s}}\right)^{-1}\left(\frac{1-f}{0.5}\right)\nonumber\\
    &&\times\left(\frac{\Delta M}{2\times10^{-2}~M_\odot}\right)\left(\frac{\theta_{\rm j}}{0.1~{\rm rad}}\right)^{-2}~\rm{erg~s^{-1}},\label{eq:Liso}
\end{eqnarray}
\normalsize
where $f$ denotes the fraction of the stripped mass lost from the system (so that $(1-f)$ is accreted onto the BH), $\eta_{\rm rad}$ is the radiative efficiency, 
and $\eta_{\rm BZ}$ is the effective efficiency of the BZ jet, incorporating the effects of BH spin and accumulated magnetic flux.
To reproduce the observed gamma-ray luminosity using the mass loss inferred from the observed timescales, a combined efficiency of $\eta_{\rm BZ}\eta_{\rm rad}\sim0.1$ is required.

Such an efficiency implies that the BH is likely rapidly spinning ($a_{\rm spin}\sim1$) and that the accretion disk is in a magnetically arrested disk state.
Combining Eqs.~(\ref{eq:pTDE}), (\ref{eq:result:a-M}), and (\ref{eq:result_e}) under the condition $a_{\rm spin}\sim1$ yields an upper limit on the BH mass of $M_{\rm BH} \lesssim 10^{6}\,M_\odot$, corresponding to a semi-major axis $a \lesssim 60\,R_\odot$.

\subsection{Precessing Orbits and Irregular Flare Intervals}
\label{sec:prec_orbit}

The irregular recurrence of the observed gamma-ray flares cannot be explained solely by repeating partial disruptions occurring at each pericenter passage. 
We therefore attribute the irregularity to precession of the jet-launching direction.

Guided by the parameters inferred in Section~\ref{sec:parameter}, we consider the BH mass $M_{\rm BH} \gtrsim 10^5\,M_\odot$, corresponding to $a \gtrsim 30\,R_\odot$ and $r_{\rm p} \lesssim 3\,r_{\rm g}$.
In this regime, the WD orbit is well approximated by a bound Kerr geodesic, which undergoes relativistic precession (see Fig.~\ref{fig:picture}(a) for an illustrative example).

To reproduce the fact that only a small number of gamma-ray flares are observed, the jet must lie outside the observer’s line of sight for most pericenter passages. 
This requires that the orbital precession angle per pericenter passage, $\theta_{\rm prec,orbit} \gtrsim 0.1~\mathrm{rad}$ ($\gtrsim 6^\circ$), exceed the jet half-opening angle adopted in Section~\ref{sec:jet}.
Assuming a rapidly spinning BH ($a_{\rm spin} \sim 1$), solutions of the Kerr geodesic equations indicate that such precession is achieved for prograde orbits with the minimum polar angle of the Kerr geodesic measured from the BH equatorial plane $\theta_{\rm min}\gtrsim0.12~\mathrm{rad}$ ($\theta_{\rm min}\gtrsim7^\circ$).
Although our geodesic calculations neglect mass loss,
this approximation is reasonable for the early passages considered, as the fractional mass loss per orbit is small.

We assume that the accretion disk formed after each pericenter passage lies in the same plane as the WD orbit at pericenter, so that the disk angular-momentum vector is aligned with that of the WD.
As a result, the disk (and the associated jet) is misaligned with respect to the BH spin axis by an angle $\theta_{\rm d}$.
Such a misaligned disk experiences a Lense–Thirring torque, causing the disk and the associated jet to precess around the BH spin axis \citep[e.g.,][]{Papaloizou1995, Fragile2007, Liska2018, Lu2024,Teboul2023}.
Based on the mass-loss rate estimated in Section~\ref{sec:parameter}, the accretion disk is expected to be in the super-Eddington regime, in which the misalignment between the disk–jet system and the BH spin axis can persist over timescales much longer than the Lense–Thirring precession timescale \citep[e.g.,][]{Franchini2016, Zanazzi2019}.
The Lense–Thirring precession timescale is given by \citet{Lu2024} as $t_{\rm prec,LT,obs} \gtrsim 200\,((1+z)/2)(r_{\rm in}/r_{\rm g})(M_{\rm BH}/10^5\,M_\odot)\,{\rm s}$, which is longer than the observed duration of a single gamma-ray flare ($\sim 100~{\rm s}$).
This implies that the jet orientation can change by a non-negligible fraction of a precession cycle during a single flare.
Therefore, the disk--jet system continues to undergo Lense--Thirring precession during an individual flare, with the jet sweeping out a cone of $\theta_{\rm d}$ around the BH spin axis.
As a consequence, the observed flare duration may depend on the instantaneous viewing geometry and precession phase, potentially leading to a moderate dispersion in flare lengths. 
Although the flares are typically of order $\sim 100$~s, the observed spread in durations is almost consistent with this picture.

Here we define $\theta_{\rm v,BH}$ as the viewing angle measured from the BH spin axis.
During the Lense--Thirring precession cycle, the instantaneous angle between the jet axis and the line of sight varies between $|\theta_{\rm v,BH}-\theta_{\rm d}|$ and $\theta_{\rm v,BH}+\theta_{\rm d}$.
If $|\theta_{\rm v,BH}-\theta_{\rm d}| > \theta_{\rm j}$, the jet is never observable (see Fig.~\ref{fig:picture}(b)).
In contrast, if $\theta_{\rm v,BH}+\theta_{\rm d} \lesssim \theta_{\rm j}$, the jet remains within the line of sight throughout the Lense--Thirring precession cycle and is therefore continuously observable (see Fig.~\ref{fig:picture}(c)).
This observability condition therefore requires that the viewing angle measured from the BH spin axis be small compared to the jet half-opening angle.

\section{Afterglow Emission}
\label{sec:afterglow}

\begin{figure}
\vspace*{5pt}
\centering
\includegraphics[width=0.45\textwidth]{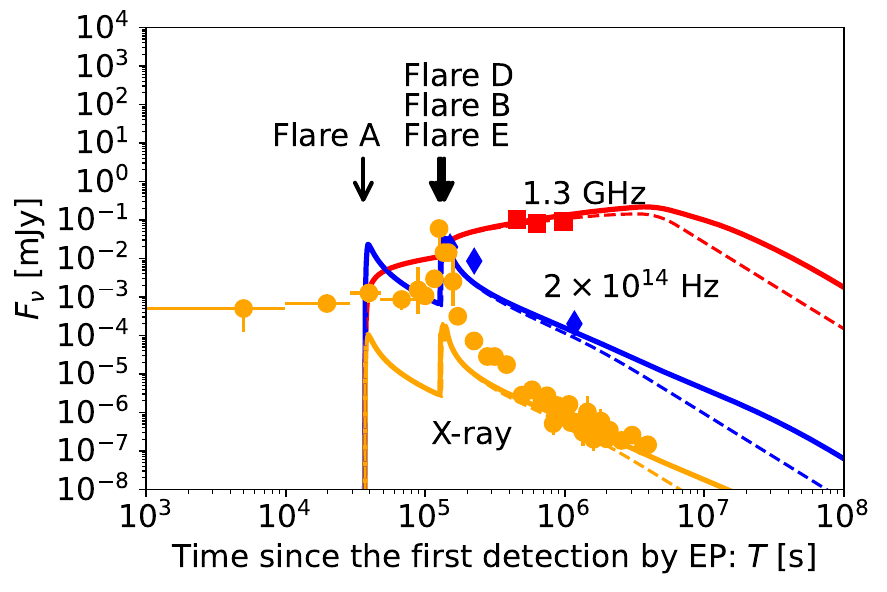}
\vspace{-0.3cm}
\caption{
Afterglow light curves in the X-ray (5~keV; orange), near-infrared ($2\times10^{14}$~Hz; blue), and radio (1.3~GHz; red) bands, compared with the observed data of GRB~250702B (X-ray: orange circles; near-infrared: blue diamonds; radio: red squares). 
The thick solid and thin dashed lines denote the predicted emission for the maximum (40 jets) and minimum (4 jets) cases, respectively.
Black arrows indicate the times of the observed gamma-ray flares.
}
\label{fig:mw}
\end{figure}
\begin{figure}
\vspace*{5pt}
\centering
\includegraphics[width=0.45\textwidth]{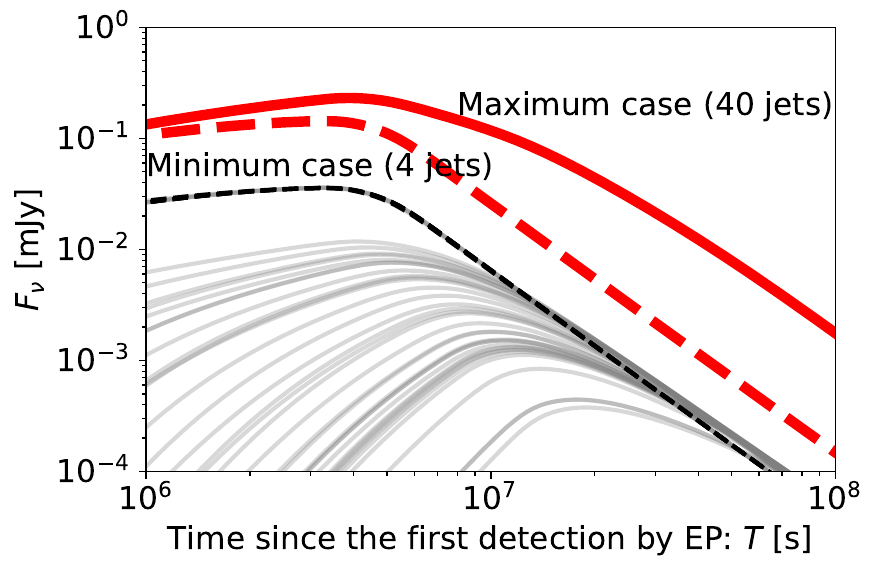}
\vspace{-0.3cm}
\caption{
Predicted radio (1.3 GHz) afterglow light curves for the maximum (40 jets; red solid line) and minimum (4 jets; red dashed line) cases.
The gray solid lines show the emission from individual jets in the maximum case, while the black dashed lines represent the emission from individual jets in the minimum case.
At $T \sim 10^{8}$ s, the maximum case is approximately an order of magnitude brighter than the minimum case.
}
\label{fig:radio}
\end{figure}

In the previous section, we consider that the WD can launch up to 40 jets over the entire sequence 
(hereafter, the maximum case).
Most of these jets are initially viewed off-axis by the observer.
As a result, the early-time emission is dominated by the four jets that happen to be launched nearly along the line of sight, while the remaining $\approx 36$ jets are initially invisible.
As the jets decelerate, relativistic beaming weakens, allowing these off-axis jets to become visible to the observer \citep[e.g.,][]{Granot2002}.
Their emission can therefore brighten at late times, particularly in the radio band \citep[e.g.,][]{Lazzati2018,Sato2021,Sato2023a}.
On the other hand, only four gamma-ray flares have been observed.
This may indicate that only four jets were launched in total (hereafter referred to as the minimum case), possibly as a consequence of disk instabilities \citep{Granot2025} and/or choked jet scenarios \citep{Lu2024}.
Consequently, the maximum and minimum cases are expected to exhibit distinct late-time radio afterglow behaviors.
In the following, we compare these two scenarios using their predicted afterglow emission.

In the maximal case, we assume that jets are launched sequentially, one every hour, corresponding to the orbital period.
For simplicity, the jet axis is assumed to remain fixed during the afterglow phase of each episode.
The total afterglow emission is computed as the sum of the radiation from 40 jets with randomly assigned viewing angles between $0$ and $\pi/2$.
Geodesic calculations show that for $\theta_{\rm min}\sim \pi/2$, the jet becomes nearly perpendicular to the BH spin axis, allowing viewing angles up to $\pi/2$ for an observer located near the spin axis.
In the minimal case, jets are launched only at the times when gamma-ray flares are observed,
and all jets are assumed to be viewed on-axis.

We compute the multiwavelength afterglow emission from a single top-hat jet using the \texttt{VegasAfterglow} code \citep{Wang2025}.
The jet is characterized by an initial half-opening angle $\theta_j$, bulk Lorentz factor $\Gamma_0$, and isotropic-equivalent kinetic energy $E_{\rm iso,K}$.
It propagates into an external medium with a power-law density profile $n(r)=A_*(r/r_0)^{-k}$, where $r_0=10^{17}\,{\rm cm}$, and produces synchrotron radiation from external shocks.
We assume a power-law electron energy distribution with index $p$.
The microphysical parameters $\epsilon_e$, $\epsilon_B$, and $f_e$ denote the fractions of internal energy in nonthermal electrons and magnetic fields, and the fraction of accelerated electrons, respectively, and are taken to be constant.
The emission is calculated for a viewing angle $\theta_{\rm v,jet}$.

We present our modeling of the multiwavelength afterglow emission in the X-ray (5~keV), near-infrared ($2\times10^{14}$\,Hz), and radio (1.3~GHz) bands, and compare the results with observations of GRB~250702B.
The X-ray data are taken from \citet{EP2025} and \citet{OConnor2025}.
We convert the observed 0.3--10~keV energy flux to a flux density at 5~keV, assuming a photon index of 1.8 during the slow-cooling phase.
The near-infrared data are from \citet{Levan2025} and \citet{Carney2025}.
We adopt a V-band extinction of $A_V=4.2$~mag \citep{JWST2025} and apply the corresponding extinction correction at $2\times10^{14}$~Hz using a standard extinction law.
The radio data are taken from \citet{Carney2025}.

In order to reproduce the observed multiwavelength afterglow light curves, we adopt the following model parameters:
$\theta_j = 0.1$~rad ($6^\circ$),
$E_{\rm iso,K} = 4\times10^{54}$~erg,
$\Gamma_0 = 100$,
$k = 1.0$,
$A_* = 0.1$,
$p = 2.6$,
$\epsilon_e = 10^{-2}$,
$\epsilon_B = 10^{-5}$, and
$f_e = 0.1$.
For a jet half-opening angle of $\theta_j\sim0.1$ rad, the collimation-corrected kinetic energy is $E_{\rm jet,K}\sim E_{\rm iso,K}\theta_0^2/2\sim2\times10^{52}$ erg per episode, which is comparable to the energy available $\eta_{\rm BZ}\Delta M c^2$.

Figure~\ref{fig:mw} presents the resulting multiwavelength emission.
In both the maximum and minimum scenarios, the early afterglow is dominated by the same four jets viewed nearly on-axis, and therefore the early X-ray and near-infrared behavior is similar in both cases.
The calculated X-ray light curve exhibits two pronounced flares between
$\sim 4.2\times10^{4}$~s and $\sim 1.2\times10^{5}$~s, each corresponding to the transition of an on-axis jet from the coasting phase to the self-similar adiabatic expansion stage \citep{Sari1997}.
The near-infrared light curve also shows two peaks, though with a different physical origin: each peak appears when the typical synchrotron frequency $\nu_m$ of the jet component crosses the observing band.
Because jets are launched at different epochs, these peaks overlap in the observer frame and produce the sequence of rapidly rising and decaying flares in both X-ray and near-infrared bands.

At later times, jets that were initially viewed off-axis in the maximum case decelerate sufficiently to become effectively on-axis \citep{Granot2002}.
For $T \gtrsim 10^{6}\,{\rm s}$, the light curves of the maximum case therefore deviate from those of the four-jet configuration, as the off-axis jets begin to satisfy the condition $1/\Gamma \sim \theta_{\rm v,jet} - \theta_{\rm j}$ and enter the observer’s line of sight, enhancing the radio emission.
This behavior is illustrated by the gray solid lines in Fig.~\ref{fig:radio}, where each radio peak corresponds to the passage of the typical synchrotron frequency $\nu_m$ through 1.3\,GHz.
By $T \sim 10^{8}$~s, all jets have entered the Newtonian phase and can be regarded as effectively on-axis.
The radio flux from the full jet population in the maximum case is then about an order of magnitude higher than that from the four-jet configuration.
Because the microphysical parameters $A_*$, $\epsilon_e$, $\epsilon_B$, and $f_e$ are identical in both scenarios, the late-time radio emission reflects the total kinetic energy,
$N_{\rm jet} E_{\rm iso,K} \theta_j^2 / 2$,
where $N_{\rm jet}$ is the number of jets.
At this epoch, $\nu_m$ lies below the radio band while the cooling frequency $\nu_c$ lies between the radio and near-infrared bands.
The flux therefore scales as $F_\nu \propto E_{\rm iso,K}^{\frac{6+10p-k(5+p)}{4(5-k)}}A_*^{\frac{19-5p}{4(5-k)}}\epsilon_B^{\frac{p+1}{4}}\epsilon_e^{p-1}f_e^{2-p}\sim E_{\rm iso,K}^{1.5}A_*^{0.4}\epsilon_B^{0.9}\epsilon_e^{1.6}f_e^{-0.6}$.
This scaling predicts the flux ratio of $(40/4)^{1.5} \sim 30$ between the maximum and minimum cases, consistent with our numerical results and yielding an order-of-magnitude difference in the radio flux at $T\sim10^{8}$~s.
At such late times, differences in jet launching epochs become negligible.

\section{Summary \& Discussion}
\label{sec:summary}

We have shown that the hierarchical timescales observed in the prompt emission of GRB~250702B
can be explained by repeating partial tidal disruptions of the WD by the BH on a highly eccentric orbit ($e\approx0.97$).
For the rapidly spinning BH with $M_{\rm BH}\sim10^{5}$--$10^{6}\,M_\odot$ and $a_{\rm spin}\sim1$, Kerr geodesics with the semi-major axis of $a\sim30$--$60\,R_\odot$ and the minimum polar angle measured from the BH equatorial plane of $\theta_{\rm min}\gtrsim0.12~\mathrm{rad}$ ($\gtrsim7^\circ$) undergo relativistic orbital precession, with the per-orbit precession angle exceeding the jet half-opening angle.
This leads to changes in the jet direction between successive pericenter passages, explaining the irregular flare intervals.
In our model, up to $\approx40$ jets are launched during the sequence of partial disruptions, but only four are initially viewed on-axis and detected as prompt gamma-ray flares.
The remaining $\approx36$ jets are initially off-axis and invisible at early times, but their cumulative contribution enhances the late-time radio afterglow by about an order of magnitude, providing a testable signature of orbital precession in the WD--IMBH system.

The last three flares appear relatively clustered. 
Since the observer is located near the BH spin axis, one possible interpretation is that the minimum polar angle, $\theta_{\min}$, could have evolved on the secular timescale, bringing the WD orbit closer to the BH equatorial plane. 
If so, this change in the orbital configuration may have led to a more stable jet direction, keeping the jet orientation nearly constant from one pericenter passage to the next.

To account for the observed gamma-ray luminosity, a magnetic field strength of $B_{\rm H} \sim 2\times10^{10}\,((1+z)/2)a_{\rm spin}(L_{\gamma, \mathrm{iso,obs}}/5\times10^{51}\,{\rm erg\,s^{-1}})^{\frac{1}{2}}(\theta_{\rm j}/0.1\,{\rm rad})(M_{\rm BH}/1\times10^5\,M_\odot)^{-1}(\kappa/0.06)^{-\frac{1}{2}}\,\mathrm{G}$ is required at the BH horizon, where $\kappa$ is a dimensionless coefficient appearing in the BZ formula \citep{Tchekhovskoy2010,Tchekhovskoy2011}.
However, the magnetic field supplied by the WD material is expected to be much weaker,
$B_{\rm adv} \sim 2\times10^{5}~(B_{\rm WD}/10^9~{\rm G})(R_{\rm WD}/0.01~R_\odot)^2(M_{\rm BH}/1\times10^5~M_\odot)^{-2}(\xi/0.1)\,\mathrm{G}$, 
where $B_{\rm WD}$ is the surface magnetic field of the WD and $\xi$ is the efficiency with which the WD magnetic field is transported to the BH horizon.
This field strength is insufficient to power the jet.
Therefore, an additional magnetic-field amplification mechanism, such as a dynamo process, would be necessary \citep[e.g.][]{Krolik2011,Tchekhovskoy2014}.

Our calculated early X-ray afterglow between $\sim 10^{4}$~s and $\sim 10^{5}$~s does not reproduce the observed behavior (see Fig.~\ref{fig:mw}).
The observed X-ray flux brightens on a timescale of about one day, peaks around the time of flare~B, and then fades on the comparable timescale.
Such day-scale variability is difficult to explain with the afterglow component.
Instead, it may arise from prompt emission associated with multiple jets viewed off-axis.
In the off-axis viewing scenario, relativistic Doppler effects reduce the observed photon energy and dilate the observed timescale, producing smoother and longer-lasting variability than an on-axis jet.

Our early near-infrared afterglow light curve also shows significant variability.
If detected by optical facilities such as the Zwicky Transient Facility or the Vera C. Rubin Observatory, such variability would provide independent evidence for repeating partial tidal disruptions.

If our model is correct, this event implies a rapidly rotating IMBH ($M_{\rm BH} \sim 10^{5}\,M_\odot$, $a_{\rm spin} \sim 1$) hosting a WD on a highly eccentric orbit.
Despite large uncertainties in their event rate \citep[e.g.,][]{MacLeod2014,Ye2023}, such systems are promising sources of low-frequency gravitational waves for space-based detectors such as the Laser Interferometer Space Antenna (LISA) \citep[e.g.,][]{LISA2023}. 
For IMBH mass, the characteristic orbital frequencies fall in the mHz band, close to the peak sensitivity of LISA.
Recent studies have highlighted BH–WD binaries \citep[e.g.,][]{Ma2025,Xuan2025}, as promising multimessenger sources in the mHz band.
Future multi-messenger observations may constrain the BH mass and spin, the orbital evolution, and enable tests of general relativity.

\begin{acknowledgements}

We thank 
Akihiro~Inoue,
Jim~Fuller, 
Jonathan~Granot,
Koki~Kin, 
Yuki~Kudoh, 
Tomoya~Suzuguchi, and
Kengo~Tomida
for valuable comments.
We also thank the anonymous referee and the editor for their helpful comments that improved the paper.
The authors are grateful to the Yukawa Institute for Theoretical Physics at Kyoto University. 
This work was completed in part during the YITP long-term workshop YITP-T-26-02, ``Multi-Messenger Astrophysics in the Dynamic Universe.''
This research was partially supported by JSPS KAKENHI Grant 
Nos.~25KJ0010 (YS), 
24K17088 (TM),
22H00130, 23H04899, and 24K00668 (KK).

\end{acknowledgements}

\bibliography{sample701}{}
\bibliographystyle{aasjournalv7}

\end{document}